\documentclass[10pt,a4paper,twocolumn, showpacs]{revtex4}
\usepackage[english]{babel}
\usepackage{graphicx,amssymb}

\begin{document}

\title[]{Experimental observation of structural crossover in binary mixtures of colloidal hard spheres}
\bigskip
\author{J\"o{}rg Baumgartl$^{1,*}$, Roel P.A. Dullens$^{1}$, Marjolein Dijkstra$^{2}$, Roland Roth$^{3}$ and Clemens Bechinger$^{1}$}
\affiliation{$^{1}$2. Physikalisches Institut, Universit\"a{}t
Stuttgart, 70550 Stuttgart, Germany\\$^{2}$Soft Condensed Matter
Group, Utrecht University, 3584 CC Utrecht, The
Netherlands\\$^{3}$Max-Planck-Institut f\"u{}r Metallforschung,
70569 Stuttgart, Germany and \\Institut f\"u{}r Theoretische und
Angewandte Physik, Universit\"a{}t Stuttgart, 70569 Stuttgart,
Germany}

\begin{abstract}
Using confocal-microscopy we investigate the structure of binary
mixtures of colloidal hard spheres with size ratio $q=0.61$. As a
function of the packing fraction of the two particle species, we
observe a marked change of the dominant wavelength in the pair
correlation function. This behavior is in excellent agreement with a
recently predicted structural crossover in such mixtures. In
addition, the repercussions of structural crossover on the
real-space structure of a binary fluid are analyzed. We suggest a
relation between crossover and the lateral extension of networks
containing only equally sized particles that are connected by
nearest neighbor bonds. This is supported by Monte-Carlo simulations
which are performed at different packing fractions and size ratios.
\end{abstract}


\maketitle

Most systems in nature and technology are mixtures of differently
sized particles. Each distinct particle size introduces another
length scale and its competition gives rise to an exceedingly rich
phenomenology in comparison with single-component systems. Already
the simplest conceivable multi-component system, i.e. a binary
mixture of hard spheres, exhibits interesting and complex behavior.
Just a few examples include entropy driven formation of binary
crystals \cite{Bartlett1992,Schofield2001,Eldrige1993}, frustrated
crystal growth \cite{deVilleneuve2005}, the Brazil nut effect
\cite{Hong2001}, glass-formation \cite{Eckert2002,Perera1999} and
entropic selectivity in external fields \cite{Roth2005}. Although
interaction potentials in atomic systems are more complex than those
of hard spheres, the principle of volume exclusion is ubiquitous and
thus always dominates the short-range order in liquids
\cite{Sastry1998}. Accordingly, hard spheres form one of the most
important and successful model systems in describing fundamental
properties of fluids and solids. It has been demonstrated that many
of their features can be directly transferred to atomic systems
where fundamental mechanisms are often obstructed by additional
material specific effects \cite{Poon1996}. Binary hard sphere
systems are fully characterized by their size ratio
$q=\sigma_{S}/\sigma_{B}$ with $\sigma_{i}$ the diameters of the
small (S) and big (B) spheres and the small and big sphere packing
fractions $\eta_{S}$, $\eta_{B}$, respectively.

The pair-correlation functions, $g_{ij}(r)$, are the central measure of structure in fluids; they describe the probability of finding a particle
of size $i$ at distance $r$ from another particle of size $j$. It is well known that all pair-correlation functions in any fluid mixture with
short-ranged interactions (not just hard spheres) exhibit the same type of asymptotic decay, which can be either purely (monotonic) exponential
or exponentially damped oscillatory (\cite{Grodon2004} and references therein). This prediction, which is valid in all dimensions, suggests that
all pair-correlation functions decay with a common wavelength and decay length in the asymptotic limit. For binary hard-sphere mixtures where
$\eta_{B}\gg\eta_{S}$ or $\eta_{S}\gg\eta_{B}$, this is rather obvious since the system is dominated by either big or small particles. The
pair-correlation functions will asymptotically oscillate with a wavelength determined either by $\sigma_{B}$ ($\eta_{B}\gg\eta_{S}$) or
$\sigma_{S}$ ($\eta_{S}\gg\eta_{B}$). Rather surprising is that the above statement is also valid for all other relative packing fractions where
the system is not dominated by particles of a single size (\cite{Grodon2004, Grodon2005}). Accordingly, in the asymptotic limit the
$(\eta_{S},\eta_{B})$ phase diagram is divided by a sharp crossover line where the decay lengths of the contributions to $g_{ij}(r)$ with the
two wavelengths become identical. Below and above this line, however, the pair-correlation function is either determined by the diameter of the
small spheres or that of the big spheres \cite{remark1}.

Despite the generic character of structural crossover and the close
relationship between structural and mechanical properties, this
effect has not been observed in experiments as the asymptotic limit
is difficult to reach in scattering experiments on atomic and
molecular liquids. However, recent calculations suggest that
structural crossover is already detectable at relatively small
distances \cite{Grodon2005}. Because colloidal particles are
directly accessible in real space, such systems provide an
opportunity to explore the structure of binary fluids and to
investigate structural crossover experimentally.

\begin{figure}[t]
\includegraphics*[width=7.2 cm]{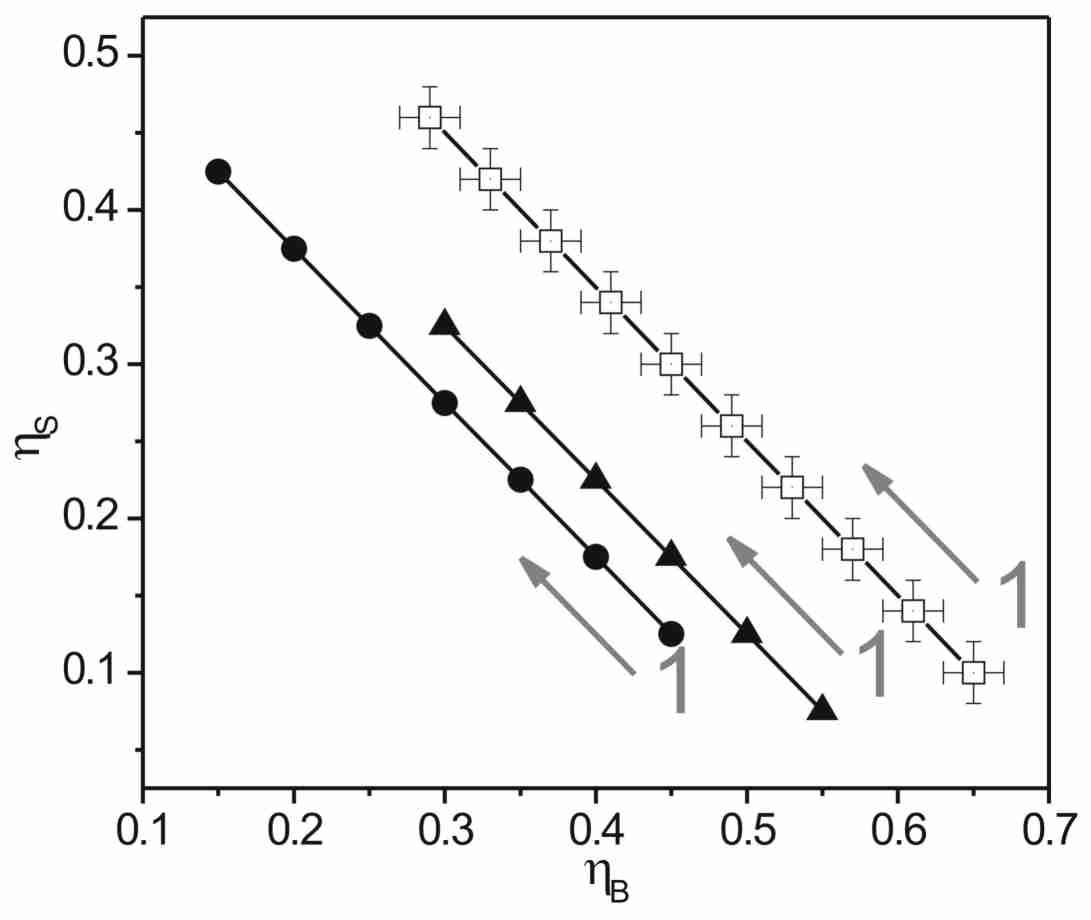}
\centering \caption{Different paths with constant total packing
fraction $\eta = \eta_{S}+\eta_{B}$ in the $(\eta_{S},\eta_{B})$-
plane. Experimental data (open symbols: $\eta=0.72$, $q=0.61$) are
sorted into ten bins. The bin size is indicated by the 'error bars'.
Closed symbols correspond to the MC-simulations ($\blacktriangle$:
$\eta=0.62$, $q=0.4$) and ($\bullet$: $\eta=0.57$, $q=0.5$). For
convenience all samples are labeled with numbers increasing in the
direction indicated by the arrows.} \label{Fig1}
\end{figure}

\begin{figure}[t]
\includegraphics*[width=8.5cm]{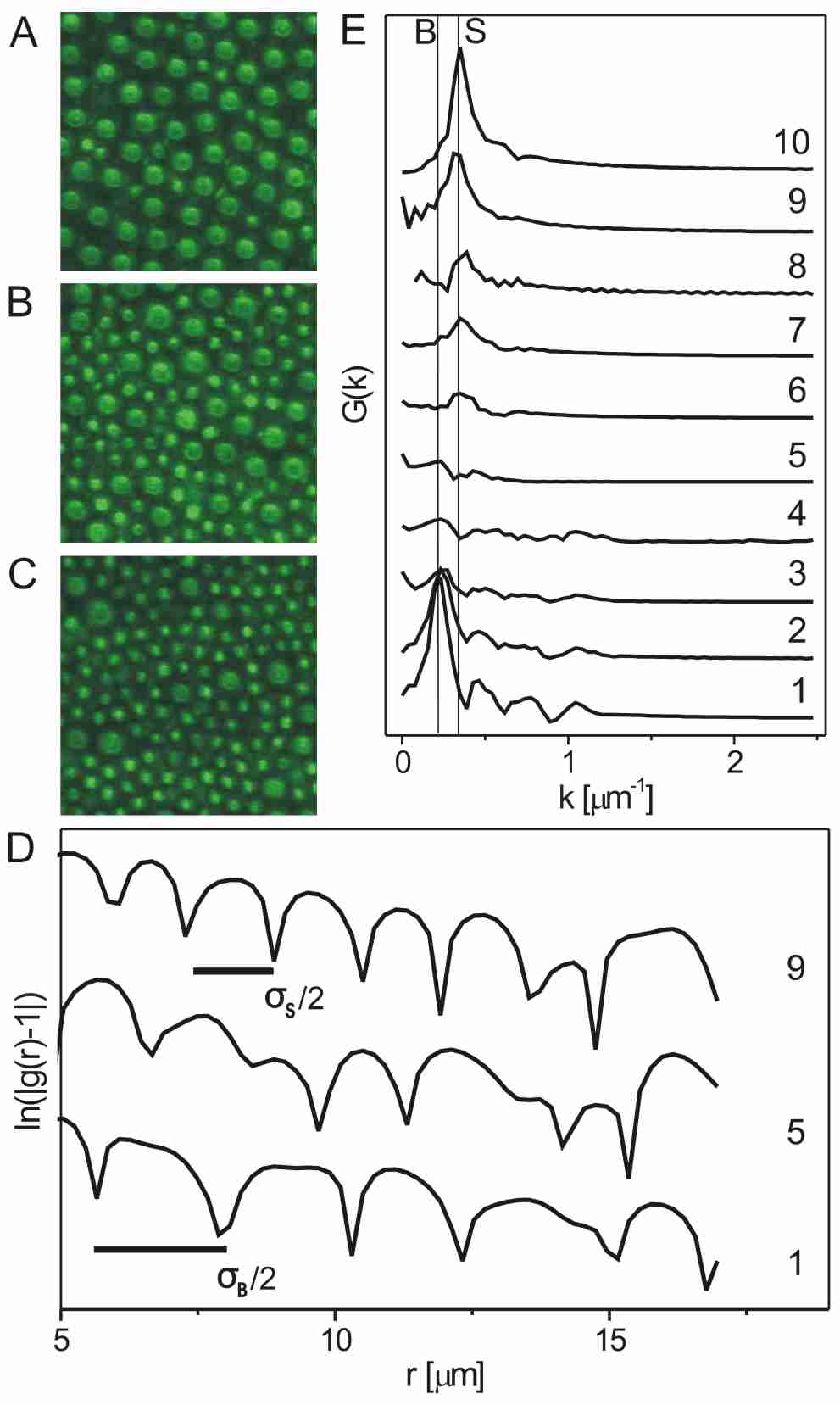}
\centering \caption{A-C) Typical snapshots of the bottom layer of a
binary mixture observed with a confocal microscope used in
reflection mode. The mixtures correspond to sample 10 (A), 5 (B) and
1 (C). The field of view is $40\times 40 \mu m^{2}$. D) Logarithmic
plot of the total correlation functions $h_{tot}(r)$ for the
experimental binary mixtures with $\eta=0.72\pm 0.04$. Correlation
functions are plotted for sample numbers 1,5 and 9 (compare
Fig.\ref{Fig1}) and are shifted in vertical direction for clarity.
The horizontal bars correspond to $\sigma_{B}/2$ and $\sigma_{S}/2$,
respectively. E) Fourier-transforms of $h_{tot}(r)$ for the
experimental data points (compare Fig. 1). Vertical lines indicate
the wave vectors $k$ corresponding to the diameters of the small (S)
and big particles (B), respectively. (color online).} \label{Fig2}
\end{figure}

\begin{figure*}[t]
\includegraphics*[width=\textwidth]{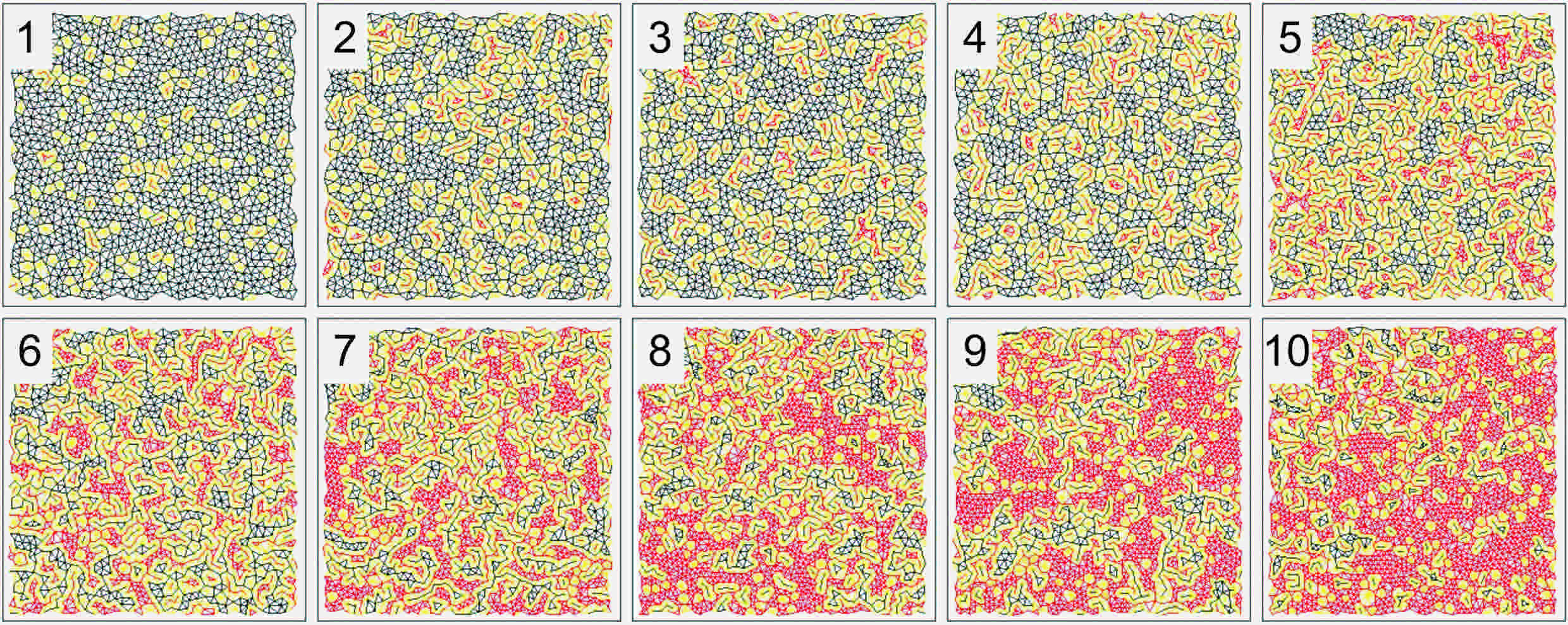}
\centering \caption{Visualization of the different bond-types as
determined by a Delaunay triangulation: big-big (black), big-small
(yellow) and small-small (red). Different plots correspond to the
sample numbers as indicated in Fig.\ref{Fig1}. The field of view is
$180\times 180 \mu m^{2}$.} \label{Fig3}
\end{figure*}

As colloidal suspension we used an aqueous binary mixture of small
melamin particles ($\sigma_{S}=2.9\mu m$) and big polystyrene
spheres ($\sigma_{B}=4.8\mu m$). Addition of salt screens residual
electrostatic interactions thus leading to an effective hard sphere
system. Since melamin has a higher density ($\rho_{M}=1.51g/cm^{3}$)
than polystyrene ($\rho_{P}=1.05g/cm^{3}$) the sedimentation
velocities are similar and, hence, we obtain a homogeneous system
after mixing. The suspension was contained in a cylindrical sample
cell with a silica bottom plate to allow optical imaging with an
inverted confocal microscope in reflection mode (Leica TCS SP2).
From the images, particle positions were obtained with digital video
microscopy \cite{Crocker1996}. Strong layering at the bottom wall
allowed us to image only the first two-dimensional bottom layer of
the three-dimensional system. We define the packing fraction as
$\eta_{i}=\pi\sigma_{i}^{2}/4$, with $\rho_{i}$ the number density
of component $i$. Variation of the relative packing fractions of the
particles was achieved by addition of small particles to a
suspension of big spheres (Fig.\ref{Fig1}). Thus, the total packing
fraction in the two-dimensional bottom layer remains constant for
all samples: $\eta=0.72$. In the following we will refer to the
different samples by the sample numbers (No.) as given in
Fig.\ref{Fig1}.

Typical snapshots of the system for different packing fractions of
big and small particles are shown in Figs.\ref{Fig2}A-C. The images
demonstrate how the structure of the bottom layer changes from being
rich in small particles (No. 1, Fig.\ref{Fig2}A) to being rich in
big particles (No. 10, Fig.\ref{Fig2}C). Fig.\ref{Fig2}B (No. 5)
corresponds to about the same number density of small and big
spheres. In order to analyze the samples for a possible structural
crossover, we calculated the pair correlation function from the
determined particle positions. To minimize statistical noise we did
not distinguish between big and small spheres. This is justified
because the crossover has been predicted to be visible in all
pair-correlation functions and thus also in any linear combination
\cite{Grodon2004,Grodon2005}. The dominating wavelength in the
oscillations is identified by computing the total correlation
function
$h_{tot}(r)=\sum_{i,j}x_{i}x_{j}h_{ij}(r)=\sum_{ij}x_{i}x_{j}[g_{ij}(r)-1]$,
with the mole fraction $x_{i}=\rho_{i}/\sum_{i}\rho_{i}$ of
component $i$ \cite{Grodon2005}. Fig.\ref{Fig2}D exemplarily shows
$\ln|h_{tot}(r)|$ for samples No. 1,5, and 9. Note that in this
representation the oscillation wavelength is halved. The correlation
functions of samples No.1 and 9 clearly oscillate with a single
wavelength, respectively, given by $\approx\sigma_{B}/2$ and
$\approx\sigma_{S}/2$. In contrast, sample 5 does not show a
dominating wavelength but an interference of different length scales
which is typical near the structural crossover. It is important to
mention, that this intermediate behavior is only observed for
samples No.5 and 6, i.e. only for about 10$\%$ of the entire range
over which $\eta_{B}$ and $\eta_{S}$ was varied. The experimentally
identified crossover-region is in excellent agreement with the
theoretically calculated value of $\eta_{S}\approx0.3$ at those size
ratios, which were determined from the decay of the pair correlation
functions calculated within density functional theory in the test
particle limit \cite{Roth2000}. Fig.2E shows the Fourier transforms
of $h_{tot}(r)$ for all samples where the rather sudden change of
the dominating wavelength is seen more clearly \cite{remark2}. At
small and high packing fractions, the correlations are clearly
dominated by frequencies corresponding to either small or large
particles (vertical lines) while around sample No.5 hardly any
dominating frequency is observed. This experimentally confirms
structural crossover as well as its occurrence at finite particle
distances.

\begin{figure}[b]
\includegraphics*[width=8.5cm]{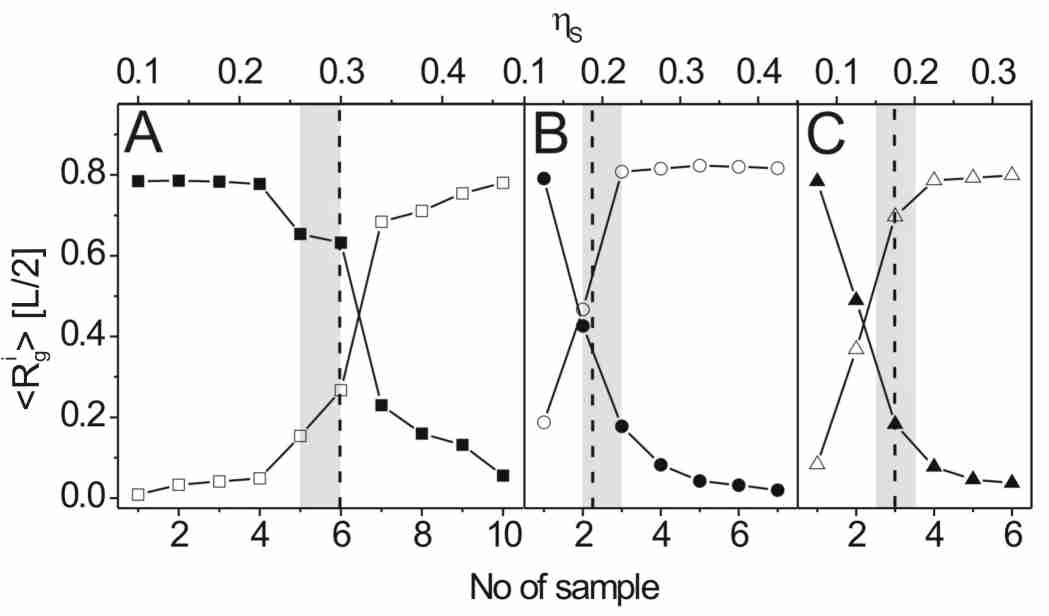}
\centering \caption{Averaged radii of gyration $\langle
R_{g}^{i}\rangle$ (normalized to $L/2$ with $L^{2}$ the size of the
field of view) of networks formed by large (solid symbols) and small
particles (open symbols) as a function of the sample number for A)
the experimental data, B) the MC-simulations at $\eta=0.57$ and
$q=0.5$ and, C) the MC-simulations at $\eta=0.62$ and $q=0.4$. The
corresponding packing fraction of small particles $\eta_{S}$ is
indicated as well. The grey area and the dashed line respectively
indicate the crossover as inferred from the correlation functions
and from density functional theory. (color online).} \label{Fig4}
\end{figure}

So far, structural crossover has been discussed in terms of pair
correlation functions, i.e. spatially averaged quantities. Since our
experiments naturally provide detailed structural information, we
investigate what the repercussions are of the structural crossover
on the real-space structure. We first subjected a Delaunay
triangulation to the set of particle centers and identified
nearest-neighbor bonds between big-big (black), big-small (yellow),
and  small-small (red) particles, respectively (see Fig.\ref{Fig3}).
As observed in Fig.\ref{Fig3}, sample 1 predominantly consists of
big-big bonds which form a large network spreading across the entire
field of view. With increasing sample No., i.e. increasing
$\eta_{S}$, the number of small-small bonds increases, which leads
to fragmentation of the big-big network into smaller, randomly
distributed patches. At large sample numbers, the role of big and
small particles is inverted and small-small bonds form a network
spanning the entire area (No.10). Having distinguished between
different bond-types, a natural and well-known measure of the
spatial extend of a network formed by $n^{i}$ particles of size $i$
at positions $\vec{x}_{k}^{i}$ ($k=1\dots n^{i}$) is given by the
radius of gyration
$R_{g}^{i}=\sqrt{\frac{1}{n^{i}}\sum_{k=1}^{n^{i}}(\vec{x}_{k}^{i}-\vec{R}_{0}^{i})^{2}}$,
with $\vec{R}_{0}^{i}$ the centroid position of the network.
Computing this quantity for all, say $N_{C}^{i}$, networks formed by
connected particles of size finally yields a weighted averaged
radius of gyration $\langle R_{g}^{i} \rangle
=\frac{1}{N^{i}}\sum_{m=1}^{N_{C}^{i}}n_{i}(m)R_{g}^{i}(m)$ where
$N^{i}$ denote the total number of particles $i$. We calculated
$\langle R_{g}^{i} \rangle$ for networks consisting of connected big
or small particles and plotted these values for our experimental
data in Fig.\ref{Fig4}A as a function of the sample number. At small
and high sample numbers the quantities saturate while a relatively
sharp transition with an intersection point occurs around sample 6.
This location is indeed in very good agreement with the crossover
transition as determined from the correlation functions in
Fig.\ref{Fig2} and density functional theory (also indicated in Fig.
\ref{Fig4}A). This suggests that the structural crossover
corresponds to a competition between the sizes of networks
consisting of connected big or small particles, respectively.

As structural crossover is also predicted for other size ratios and
packing fractions, we use Monte-Carlo (MC) simulations to test our
findings for more dilute systems with size ratios $q=0.5$ and
$q=0.4$. The corresponding paths through the phase diagram (see
closed symbols in Fig.\ref{Fig1}) were obtained from 2-dimensional
simulations with a fixed number of particles of about $0<N<3000$ for
both species and box areas of about $1500\sigma_{B}^{2}$ employing
periodic boundary conditions. From the configurational snapshots we
first determined the region of crossover by analyzing $h_{tot}(r)$
(the correlation functions are  sampled using $10^{4}$ MC cycles per
particle). Then, we performed the above described Delaunay
triangulation to calculate $\langle R_{g}^{i}\rangle$ for networks
of connected big or small particles, respectively. The corresponding
radii of gyration are plotted in Fig.\ref{Fig4}B and C and show a
similar behavior as in the experiment. Again, the intersection
points are consistent with the crossover region as inferred from the
correlation functions and DFT calculations. Note that the crossover
region sensitively depends on the size ratio and packing fractions.
Both the experiment and Monte-Carlo simulations show that structural
crossover is accompanied by a pronounced change in the typical size
of networks consisting of connected big and small particles. By
introducing small particles into a system of big spheres,
connections between big particles are broken and, at the same time,
connections between small particles are made. This sensitively
affects the typical size of networks containing connected,
equally-sized particles and thereby the chance of finding another
particle with the same size at a relatively large distance.
Consequently, the change from $\langle R_{g}^{B}\rangle > \langle
R_{g}^{S}\rangle$ to $\langle R_{g}^{S}\rangle > \langle
R_{g}^{B}\rangle$ (and vice versa) provides a simple real-space
argument why the oscillation wavelength of the $g_{ij}(r)$ in the
asymptotic limit is either set by $\sigma_{B}$ or $\sigma_{S}$.

We have experimentally demonstrated the structural crossover in a
binary colloidal hard sphere system. Furthermore, we show that
structural crossover is strongly coupled to the size of networks
containing connected equally-sized particles only. Going across the
structural crossover, the size ratio of such networks comprised by
either connected big or small particles is reversed. We believe this
real-space configurational picture of structural crossover is not
just applicable to binary hard spheres, as structural crossover is a
generic feature of mixtures with competing length scales. Moreover,
it shows interesting similarities with force chains in granular
matter \cite{OHern2001} and glassy systems \cite{Eckert2002,
Perera1999, Hoffmann2006} of dissimilar sized particles. Therefore,
our finding may help to gain more insight into structure-related
properties in binary systems at an universal level.

\noindent $^*$Electronic address:
j.baumgartl@physik.uni-stuttgart.de

\bibliographystyle{apsrev}

\begin{thebibliography}{18}
\bibitem[{\citenamefont{remark}(1)}]{remark1}
\bibinfo{journal}{For very asymmetric size ratios, i.e. $q<0.3$, there can
be additional regions in which oscillations at intermediate
wavelength can be observed.}
\bibitem[{\citenamefont{remark}(2)}]{remark2}
\bibinfo{journal}{In two dimensions for radial symmetric functions the
Fouriertransform becomes a Besseltransform. However, for the identification
of the dominant wavelength the usual Fouriertransform, which is numerically
easier to handle, predicts equivalent results}
\end{thebibliography}

\end{document}